\documentstyle[aps,preprint,tighten,epsfig,floats]{revtex}
\newcommand{\lsim}
 {\ \raise.37ex\hbox{$<$}\kern-0.75em\lower.7ex\hbox{$\sim$}\ }
\newcommand{\gsim}
 {\ \raise.37ex\hbox{$>$}\kern-0.75em\lower.7ex\hbox{$\sim$}\ }
\begin{document}
\sloppy
\title{New aspects of Verway transition in magnetite}
\author{Hitoshi Seo$^1$, Masao Ogata$^2$, and Hidetoshi Fukuyama$^3$}
\address{$^1$ Correlated Electron Research Center (CERC), \\
National Institute of Advanced Industrial Science and
Technology, 
Ibaraki 305-8562, Japan}
\address{$^2$ Department of Physics, University of Tokyo, 
Tokyo 113-0033, Japan}
\address{$^3$ Institute for Solid State Physics, University of Tokyo, 
Chiba 277-8581, Japan}
\date{\today}
\maketitle

\begin{abstract}
A new mechanism of the Verway transition in magnetite (Fe$_3$O$_4$), 
which has been argued to be a charge ordering transition so far, 
is proposed. 
Based on the mean field calculations 
for the three band model of spinless fermions 
appropriate for the $d$ electrons of 
the Fe ions on the $B$ sites, 
it is indicated that the 
phase transition should be the bond dimerization 
due to the cooperative effects of 
strong electronic correlation 
and electron-phonon interaction. 
The results show that 
the ferro-orbital ordered state is stabilized 
in a wide temperature range 
due to the strong on-site Coulomb interaction 
between different $t_{2g}$ orbitals 
resulting in an effectively one-dimensional 
electronic state, 
which leads the system toward an insulating state 
by the Peierls lattice distortion  
with the period of two Fe($B$)~ions, i.e., bond dimerization. 
Furthermore, 
it is found that 
the interplay between such lattice distortion 
in the Fe($B$)~ions 
and the lattice elastic energy 
of the Fe($B$)-O as well as the Fe($A$)-O bonds 
gives rise to a competition between 
two different three-dimensional patterns for the bond dimerization, 
and can stabilize a complicated one
with a large unit cell size. 
The results are compared with the known experimental facts.  
\smallskip

\end{abstract} 

\newpage

\section{Introduction}\label{intro}

\begin{figure}
 \centerline{\epsfig{file=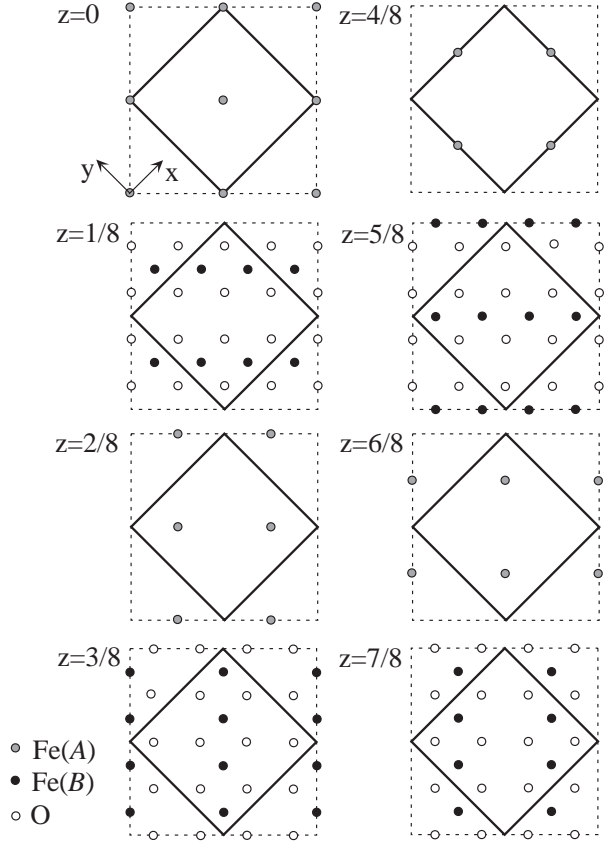,width=7.9cm}}
 \vspace{2mm}
 \caption{Cubic spinel structure of Fe$_3$O$_4$. 
 The thick square shows the cubic unit cell 
 with a lattice constant $a$ in the $xy$ plane 
 while the $z$ coordinates are indicated by the fraction of $a$.
 The unit cell in the $xy$ plane for $T < T_{\rm V}$ is shown 
 by the dotted line.}
 \label{structure}
\end{figure}

The nature of the Verway transition in magnetite (Fe$_3$O$_4$) 
has not been clarified yet, 
in spite of intensive studies 
from very early days~\cite{reviews}. 
Fe$_3$O$_4$ forms the cubic spinel structure, 
as shown in Fig.~\ref{structure}, 
where one-third of the Fe~ions occupy 
the $A$~sites, tetrahedrally coordinated by 4 oxygen ions, 
while the remaining two-third are octahedrally surrounded 
by 6 oxygen ions, which are the $B$~sites.
One can see that Fe($B$)$_2$O$_4$~layers and Fe($A$)~layers 
are stacked alternatively, 
while the $z$ axis can be taken as any of the 
three cubic crystal axes. 
Although the primitive cell is a rhombohedral parallelepiped 
containing 2 formula units of Fe$_3$O$_4$, 
the unit cell is conveniently chosen as the cubic one,
as shown in Fig.~\ref{structure}, 
containing 8 formula units, 
to which we also refer in this paper. 

The Fe($A$)~ions are trivalent 
while the Fe($B$) ions are mixed valent 
with formal average valence Fe($B$)$^{2.5+}$. 
Below $T_{\rm N}=$ 858 K, 
the magnetic moments of the Fe ions are ferrimagnetically ordered, 
where the $A$~sites and the $B$~sites have opposite spin directions, 
with $d$-orbital occupation 
represented as 
$(t_{2g\uparrow})^3(e_{g\uparrow})^2$ 
and $(t_{2g\downarrow})^3(e_{g\downarrow})^2(t_{2g\uparrow})^{0.5}$, 
respectively.
Well below $T_{\rm N}$, 
an abrupt increase in the resistivity takes place at $T_{\rm V}\simeq$~120~K, 
which is now called after Verway, 
who has discovered it~\cite{Verway} 
and proposed that the phenomena is due to 
the ordering of equal number of
Fe$^{2+}$ and Fe$^{3+}$ on the $B$~sites, 
i.e., charge ordering~(CO) among 
the ``extra'' $t_{2g\uparrow}$~electrons~\cite{Verway2}. 

As for theoretical studies on this phenomena, 
the electronic properties of the 
spin polarized $t_{2g\uparrow}$~electrons 
have been investigated frequently by 
the single band spinless fermion model on the $B$ sites 
originally proposed by Cullen and Callen~\cite{CandC,Ihle}. 
Such a picture of spinless fermion
has been proved to be valid 
by band structure calculations~\cite{Yanase,Zhang,Anisimov}, 
where the majority spin~($\downarrow$) band of the Fe($B$) ions 
is shown to be fully occupied and 
the Fermi energy crosses the minority spin~($\uparrow$) band.
In the model of Cullen and Callen 
the triple degeneracy of the $t_{2g}$~orbitals is neglected 
so that the band is half-filled, 
corresponding to half a charge per site as in the actual compound. 
This model is investigated within the Hartree approximation 
where the ground state shows a phase transition 
from a metallic state to a CO state 
as the intersite Coulomb interaction $V_{ij}$ is increased~\cite{CandC}, 
providing the most naive picture for the CO state 
in the magnetite. 

However, 
no CO pattern so far considered has been successful in presenting 
a satisfactory interpretation 
for the known experiments, 
including the one originally proposed by Verway~\cite{Verway2} 
and other more complicated patterns~\cite{Mizoguchi,Zuo}.
Besides,
the expected short range fluctuation of the CO 
due to the frustration in $V_{ij}$~\cite{Anderson},
arising from the fact that the network connecting 
the Fe($B$)~ions forms a coupled tetrahedra system, 
is not found either in neutron scattering~\cite{Fujii} 
or in resonant X-ray scattering measurements~\cite{Garcia}.
Furthermore, 
the recent 
NMR~\cite{Novak} and X-ray anomalous scattering~\cite{Garcia2} 
experiments 
for $T<T_{\rm V}$ even cast doubts on the existence of CO. 
Todo {\it et al.}~\cite{Todo}, 
based on these suggestions 
and their finding of 
pressure-induced metallic ground state above 8 GPa, 
proposed that the low temperature phase below $T_{\rm V}$ 
may be a kind of ``Mott insulator'' where the $B$ sites 
are forming dimers due to orbital ordering (OO). 

This idea of the Mott insulating state resembles to that 
in low-dimensional quarter-filled organic conductors 
theoretically studied by Kino and the present authors~\cite{Fukuyama},
where the fact that there exists a carrier per 2 sites 
is the same as that for the $B$~sites in Fe$_3$O$_4$. 
In such organic conductors, 
the Mott insulating state can be understood from a view point 
that each charge localizes in every two sites, 
i.e., dimer. 
The dimerization is either due to the anisotropy in the transfer integrals 
from the lattice structure~\cite{Kino}, 
or due to the spontaneous formation of 
bond dimerization~(BD) by the 
electron-phonon interaction 
where the one-dimensionality (1D) plays a crucial role~\cite{bonddimer}.
\begin{figure}[b]
 \centerline{\epsfig{file=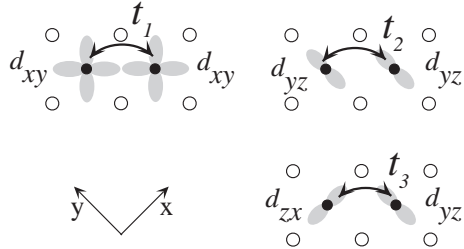,width=6.5cm}}
 \vspace{2mm}
 \caption{A schematic representation of the transfer integrals 
 between neighboring $B$ site pairs in the $xy$ plane
 with different orbital occupations.
 The simultaneous cyclic substitution of the orbital indices 
 and the axes of coordinates, 
 $x \rightarrow y, y \rightarrow z, z \rightarrow x$ 
 or $x \rightarrow z, y \rightarrow x, z \rightarrow y$, 
 also provides the three transfer integrals, 
 while for the other configurations $t_{ij}^{\mu\nu}$=0.}
 \label{transfer}
\end{figure}

In this paper, 
we will show that this latter type of insulating state with BD 
can actually emerge in Fe$_3$O$_4$, 
and propose it to be the mechanism of the Verway transition 
in this compound. 
We will discuss in Sec.~\ref{secBsites} 
that the strong correlation among electrons 
stabilizes an OO state with effectively 1D electronic structure, 
and additional electron-phonon interaction 
can give rise to the BD state. 
Moreover, 
discussions in Sec.~\ref{secBDpattern} 
on the elastic lattice energy of the whole system 
will lead to the three-dimensional pattern of this BD 
expected below $T_{\rm V}$. 
The relevance of our proposal to the experimental facts 
is discussed in Sec.~\ref{secexp}, 
and the conclusion is given in Sec.~\ref{secconcl}.


\section{Bond Dimerization on $B$ sites}\label{secBsites}

\subsection{Three band model}\label{subsec3band}

We start with the model of Mishra~{\it et~al.}~\cite{Mishra},
who have extended 
the spinless fermion model of Cullen and Callen~\cite{CandC} 
by including 
the triple degeneracy of the $t_{2g}$~orbitals 
appropriate for the $B$~sites in Fe$_3$O$_4$.
The Hamiltonian is written as 
\begin{eqnarray}
H=\sum_{\langle ij \rangle}\sum_{\mu\nu} t_{ij}^{\mu\nu}& c^{\dagger}_{i\mu} &
    c_{j\nu}+\sum_{i}\sum_{ \mu\ne\nu } U n_{i\mu}n_{i\nu} \nonumber\\
    &+& \sum_{\langle ij \rangle } V n_{i}n_{j},
\label{eqn:H}
\end{eqnarray}
where 
$c^{\dagger}_{i\mu}$ ($c_{i\mu}$) and $n_{i\mu}$ denote the 
creation (annihilation) and the number operator 
of the electron at $i$th site of orbital $\mu$, 
respectively, 
where the orbital indices take $xy$, $yz$ or $zx$~\cite{note}. 
$n_i$ is the number operator for each site, 
i.e., $n_i = n_{ixy}+n_{iyz}+n_{izx}$, 
and $\langle ij \rangle$ denotes the nearest-neighbor site pair 
along the coupled Fe($B$) ion tetrahedra network. 
$U$ and $V$ are the on-site Coulomb energy between different orbitals 
(note that the same orbital cannot be doubly occupied) 
and the nearest-neighbor Coulomb energy, 
respectively.
By using the transfer integrals for $\langle ij \rangle$ pairs 
calculated in Ref.~\ref{Zhang}, 
$t_{dd\sigma} = -0.41$~eV, $t_{dd\pi} = 0.054$~eV
and $t_{dd\delta}=0.122$~eV, 
the three kinds of transfer integrals, $t_{ij}^{\mu\nu}$, 
with 
different configurations of orbital occupations, 
as shown in Fig.~\ref{transfer},
can be estimated~\cite{Harrison}
as $t_1=3t_{dd\sigma}/4+t_{dd\delta}/4=-0.278$~eV, 
$t_2=t_{dd\pi}/2+t_{dd\delta}/2=0.085$~eV
and $t_3=t_{dd\sigma}/2-t_{dd\delta}/2=-0.035$~eV, 
where the notations of $t_{1\sim3}$ are given in Fig.~\ref{transfer}.

Mishra {\it et~al.} treated 
the Coulomb interaction terms, $U$ and $V$, 
by means of Hartree mean field approximation,
\begin{eqnarray}
n_{i\mu}n_{j\nu} \rightarrow 
\left<n_{i\mu}\right>n_{j\nu}
+n_{i\mu}\left<n_{j\nu}\right>
-\left<n_{i\mu}\right>\left<n_{j\nu}\right>,
\label{eqn:MF}
\end{eqnarray}
and determined the ground state of the system. 
Their results for the case with the fixed value of $U~=~4.0$~eV, 
relevant for magnetite~\cite{Zhang}, 
showed that the CO state is stabilized in the ground state
for $V~>~V_{\rm c}=0.38$~eV~\cite{note2} 
while the metallic state is for $V~<~V_{\rm c}$, 
qualitatively the same as in the one-band model~\cite{CandC}, 
and they concluded that the former state is relevant for the magnetite. 

Although it is not emphasized in their paper, 
every site is almost fully 
occupied by the same orbital state, 
namely, ferro-OO is realized in both of these states, 
due to the large value of $U$. 
The three $d_{xy}$-, $d_{yx}$- and $d_{yz}$-OO states 
are energetically degenerate, 
and the electronic structure becomes 1D 
once one of the OO state is realized, 
which is particular to the spinel structure. 
For example the $d_{xy}$-OO state 
has 1D arrays of sites connected by the largest transfer integral $t_1$, 
which are along the $[110]$ direction and
the $[1{\bar 1}0]$ direction for the $xy$ planes with $z$ coordinates 
$z=(4n+1)/8$ and $z=(4n+3)/8$, respectively (see Fig.~\ref{structure}).  

In the metallic state 
with the presence of this $d_{xy}$-OO, 
the expectation values of the charge density for all sites are 
$\langle n_{ixy} \rangle \simeq 0.5$ and 
$\langle n_{iyz} \rangle = \langle n_{izx} \rangle \simeq 0$. 
As for the CO state, 
i.e., the coexistent state of OO and CO, 
the planes with 1D arrays of $\langle n_{ixy}\rangle \simeq 0.5+ \delta$ and 
those with  $\langle n_{ixy}\rangle \simeq 0.5 - \delta$, 
corresponding to Fe$^{2+}$ and Fe$^{3+}$, respectively, 
are stacked alternatively along the $z$ direction, 
where the CO pattern is the one proposed by Verway. 
Here $\delta$ is the amount of charge disproportionation, 
which rapidly increases as the value of $V$ is increased from $V_{\rm c}$, 
and reaches around 0.9 for $V=1.0$~eV. 
The other two $d_{yz}$ and $d_{zx}$ orbitals 
are almost empty also in the CO state. 

In contrast to their conclusion that this CO state 
is realized in magnetite, 
the recent experimental proposals of the absence of CO 
mentioned in Sec.~\ref{intro} 
suggest that the former {\it OO metallic} state should be 
relevant to the actual system, 
and that the origin of insulator is other than CO. 
The destabilization of the CO state 
may be due to the screening of the long range Coulomb interaction 
and/or the effect of frustration among $V_{ij}$ 
mentioned above~\cite{Anderson}. 
Actually, 
the melting of CO due to frustration among $V_{ij}$ 
has been demonstrated theoretically in 1D systems~\cite{SeoPr}, 
though we will not discuss such possibility in magnetite 
further in this paper. 
Here we will concentrate in the following on 
how the OO metallic state shows 
instability toward an insulating state 
and its possible consequence on the lattice structure 
in this insulating phase. 

\subsection{Peierls instability in orbital ordered state}\label{subsec1D}

In the following we consider the case of the $d_{xy}$-OO metallic state. 
The properties of this ferro-OO state can be extracted by an effective 
noninteracting 1D spinless fermion system, 
$H_{\rm 1D}=\sum_{l} t_1 (c^{\dagger}_l c_{l+1}+h.c.)$,
%
%
with a half-filled band, 
$l$ being the site index along the 1D directions $[110]$ or $[1{\bar 1}0]$.
This model should be valid for 
$U \gg t_1 \gg t_2,t_3$ in eq. (\ref{eqn:H}) with $V_{ij}=0$. 
It is well known that such a half-filled 1D band 
has the Peierls instability 
with the wavelength of two times the inter-atomic
distance~\cite{bonddimer}, 
i.e., {\it instability toward the BD state} 
with alternating atomic displacements along the 1D direction. 
Actually the system 
can be mapped onto the $S=1/2$ XY chain 
via Jordan-Wigner transformation, 
so that our metallic state here 
correspond to the spin liquid state, 
which shows instability toward 
the spin-Peierls state, 
that is, the BD state in spin system. 
Thus, 
once the electron-phonon interaction is present,
the OO state in magnetite will undergo a BD transition 
to gain kinetic energy by making a gap at the band center, 
resulting in an {\it insulating} state.  
We propose this scenario to be the origin 
of the insulating ground state of magnetite,
Fe$_3$O$_4$. 
Such a state will be realized 
if it is not destroyed by the inter-chain interactions $t_2$ and $t_3$, 
which is investigated in the next subsection. 

We should note that this Peierls instability is 
different from the usual one in 
the 1D electron system of weak coupling, 
since the assumption of the spinless fermion here 
is due to the ferrimagnetic spin ordering 
realized in the limit of infinite value of on-site intra-orbital Coulomb 
interaction.
Thus one may say that the BD state here 
is rather analogous to the Mott insulator in 
quarter-filled compounds noted 
in Sec.~\ref{intro}~\cite{Fukuyama,Kino,bonddimer}, 
where the origin is also the interplay between 
the strong on-site Coulomb interaction 
and the dimerization.

\subsection{Peierls-Hubbard model}\label{subsecPH}

To investigate the stability of the BD state 
in the actual three-dimensional system,
we include the lattice degree of freedom 
by adding the Peierls-type coupling to Eq.~(\ref{eqn:H}), 
i.e., treat the Peierls-Hubbard model for magnetite. 
It is expressed as 
\begin{eqnarray}
H_{\rm PH}=\sum_{<ij>}\sum_{\mu\nu} t_{ij}^{\mu\nu}(1+u_{ij}) 
    & c^{\dagger}_{i\mu} &
    c_{j\nu}+\sum_{i}\sum_{ \mu\ne\nu } U n_{i\mu}n_{i\nu} \nonumber\\
    &+& \sum_i \frac{1}{2}K u_{ij} ^2,
\label{eqn:H_P-H}
\end{eqnarray}
\begin{figure}
 \centerline{\epsfig{file=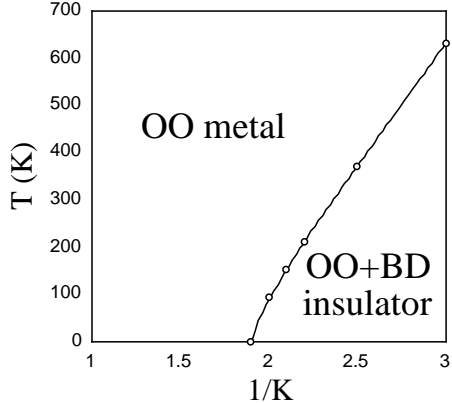,width=6cm}}
 \caption{Mean field phase diagram for the three-band 
 Peierls-Hubbard model for the $B$ sites, 
 on the plane of temperature, $T$, and the inverse of the lattice elastic
 coupling constant, $1/K$. 
 OO and BD represent orbital-ordering and bond dimerization, respectively.}
 \label{phasediagram}
\end{figure}
\begin{figure}
 \centerline{\epsfig{file=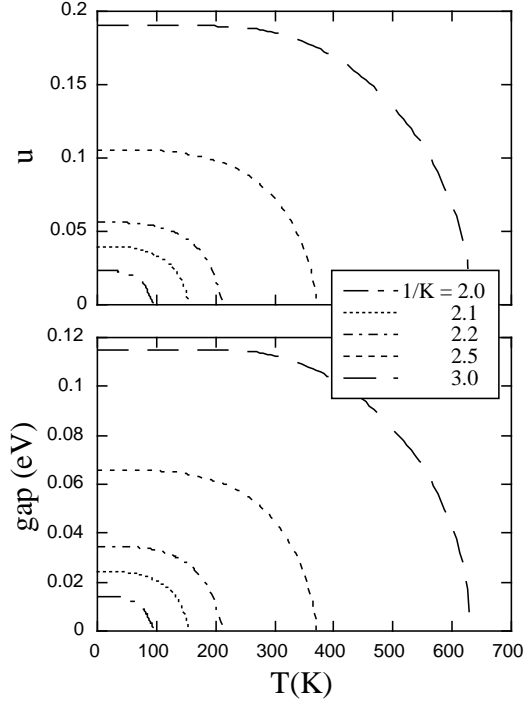,width=7cm}}
 \caption{Temperature dependences of 
 (a) the degree of BD,~$u$, and (b) the band gap 
 for several values of $1/K$.}
 \label{ugapT}
\end{figure}
where $u_{ij}$ and $K$ are the lattice distortion 
between nearest-neighbor site pair $\langle ij \rangle$ 
and the coupling constant for the elastic energy, respectively. 
The intersite Coulomb interaction term is neglected here
since it is not relevant in our discussion 
as mentioned in Sec.~\ref{subsec3band}. 
The on-site Coulomb interaction term, $U$, is treated 
within the mean field approximation, as in eq.~(\ref{eqn:MF}), 
and self-consistent solutions are obtained. 
We restrict ourselves to the case of 
$u_{ij}=(-1)^l u$ associated with the three kinds of transfer integrals 
along the [110] and [1${\bar 1}$0] directions 
and $u_{ij}=0$ for the other directions.
This provides the BD state, $u$ being the degree of BD, 
and the value of $u$ is determined for each choice of parameters 
so as to minimize the energy. 
We note that 
the calculated energy does not depend on different BD patterns 
since the effect of the lattice distortion 
on the lattice elastic energy 
and on the transfer integrals 
is restricted in each chains. 

The results show that the BD state can actually be stabilized 
when $1/K$ exceeds a critical value. 
In Fig.~\ref{phasediagram}, 
the obtained phase diagram within 
the finite temperature mean field approximation 
is shown for the case of fixed $U=4.0$~eV, 
the same as in the calculation 
of Mishra~{\it et~al.}~\cite{Mishra} mentioned in Sec.~\ref{subsec3band}. 
The ground state is 
OO metallic with $u=0$ below $1/K < 1.9$ eV$^{-1}$, 
whereas the OO state with $u \ne 0$, 
i.e., the BD state, is stabilized for $1/K  > 1.9$ eV$^{-1}$, 
where the system is insulating. 
Below $T_{\rm OO}\simeq$ 6000~K, 
which is the temperature range 
shown in Fig.~\ref{phasediagram}, 
the ferro-OO is present, 
and the para-orbital state is stabilized only above $T_{\rm OO}$. 

The optimized value of $u$ and the band gap 
in the BD state 
are plotted as a function of $T$ 
in Figs.~\ref{ugapT}(a) and \ref{ugapT}(b), respectively.
It can be seen that the lattice dimerization 
and the band gap decrease 
as the temperature is increased and vanish
continuously at $T=T_{\rm c}$, 
showing a second order insulator-to-metal phase transition. 
On the other hand, 
the charge density for each orbital state 
does not change noticeably from 
$\langle n_{ixy}\rangle \simeq 0.5$ and 
$\langle n_{iyz}\rangle=\langle n_{izx}\rangle \simeq 0$, 
for all sites, 
in the temperature range we are concerned.  
In other words, 
the OO is not affected through the metal-insulator transition. 


\section{Bond dimerization pattern}\label{secBDpattern}

The three-dimensional BD pattern in the actual compound 
cannot be determined by the preceding calculation alone, 
since the calculated energy does not depend 
on the inter-chain configuration, 
as mentioned above. 
To discuss theoretically the stability of different BD states, 
the lattice elastic energy should be the most important factor
since the BD produces large lattice distortion 
so that it moderately affects the lattice energy. 
The influence of the BD on the the inter-chain transfer integrals 
would be small 
so the difference in the kinetic energy 
between different BD states 
is neglected in the following discussion. 

The lattice elastic energy in Fe$_3$O$_4$, $\mathcal{E}_{\rm lat}$, 
can be estimated by 
the sum of 
$\frac{K}{2}(\Delta u)^2$ for all the nearest neighbor Fe-O bonds,
where $\Delta u$ is the deviation of the Fe-O distance from 
that in the equilibrium position above $T_{\rm V}$ 
and $K$ is the elastic constant. 
The value of $K$ should take 
common values $K_{\rm A}$ for the Fe($A$)-O bonds 
and $K_{\rm B}$ for the Fe($B$)-O bonds, 
since all the Fe($A$)-O bonds as well as the Fe($B$)-O bonds 
are respectively crystallographically equivalent above $T_{\rm V}$. 
Thus $\mathcal{E}_{\rm lat}$ 
can be expressed as $\mathcal{E}_{\rm Alat}$ + $\mathcal{E}_{\rm Blat}$, 
for the Fe($A$)-O and the Fe($B$)-O bonds, respectively, 
where 
\begin{eqnarray} 
{\mathcal E}_{\rm Alat} = 
\sum \frac{K_{\rm A}}{2}(\Delta u_{\rm A-O})^2,\label{Alat}\\ 
{\mathcal E}_{\rm Blat} =
\sum \frac{K_{\rm B}}{2}(\Delta u_{\rm B-O})^2, \label{Blat}
\end{eqnarray}
where 
$\Delta u_{\rm A-O}$ and $\Delta u_{\rm B-O}$ 
are $\Delta u$ for the Fe($A$)-O and Fe($B$)-O bonds, respectively.
\begin{figure}
 \centerline{\epsfig{file=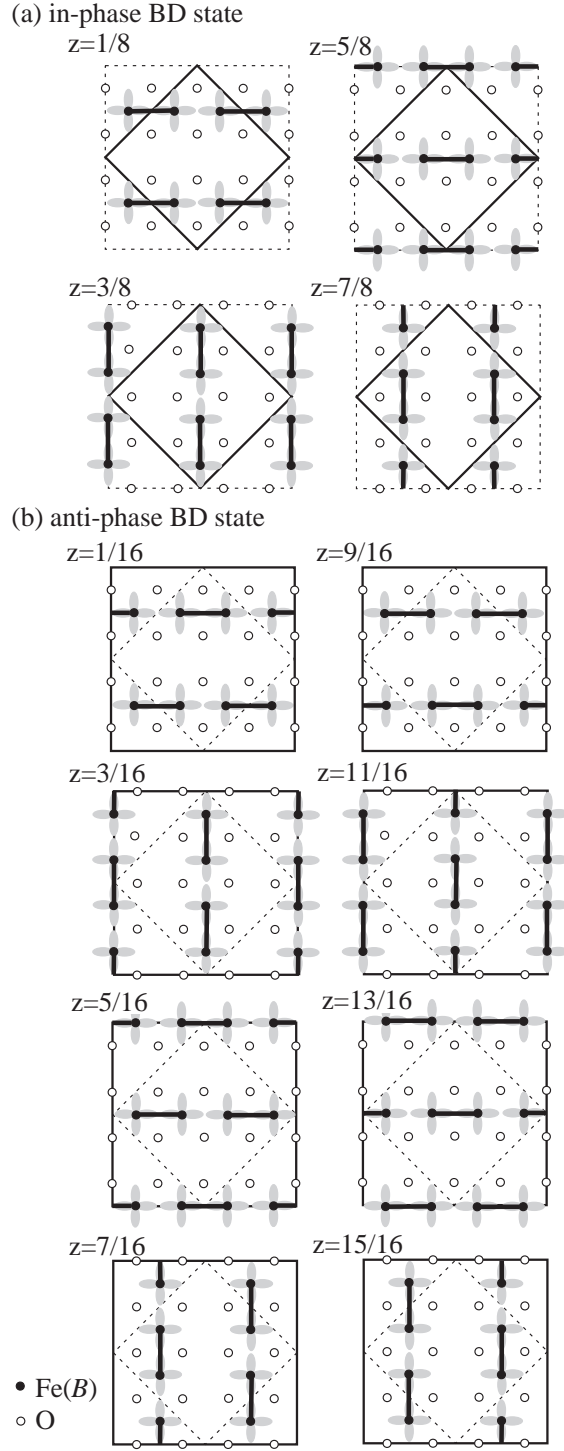,width=7.5cm}}
 \vspace{2mm}
 \caption{Schematic representation of 
 (a) the in-phase BD state and (b) the anti-phase BD state 
 both coexisting with the $d_{xy}$-OO. 
 Only the Fe($B$)$_2$O$_4$ layers are shown, 
 where the thick Fe($B$)-Fe($B$) bonds represent the dimers. 
 The thick squares show the unit cell in the $xy$ plane 
 and the $z$ coordinates indicate the fraction of 
 unit cell size along the $z$ direction. 
 The unit cell remains unchanged from the cubic one above $T_{\rm V}$ 
 shown in Fig.~\ref{structure} for the in-phase BD pattern, 
 while it becomes $\sqrt{2} \times \sqrt{2} \times 2$ of that 
 for the anti-phase BD pattern.}
 \label{oodimer3}
\end{figure}

Then, as will be explained in detail later,
there are two candidates for the BD states 
costing much less elastic energy than the others, 
as schematically shown in Fig. \ref{oodimer3}(a) and (b). 
In the former the dimerization pattern between adjacent 1D chains 
in the $xy$ plane is in-phase, 
thus we call it the in-phase BD state, 
where the unit cell is unchanged from the cubic unit cell 
above $T_{\rm V}$. 
On the other hand, 
in the latter pattern the BD is anti-phase, 
which is called anti-phase BD state in the following. 
Here, the unit cell becomes large 
as $\sqrt{2} \times \sqrt{2} \times 2$ 
of the cubic one, as shown in the Figure. 
We will see below that 
the competition between two BD states arises 
depending on the relative value of $K_{\rm A}$ and $K_{\rm B}$, 
and our proposal is that 
the anti-phase BD pattern is realized in 
the actual compound. 

\begin{figure}
 \centerline{\epsfig{file=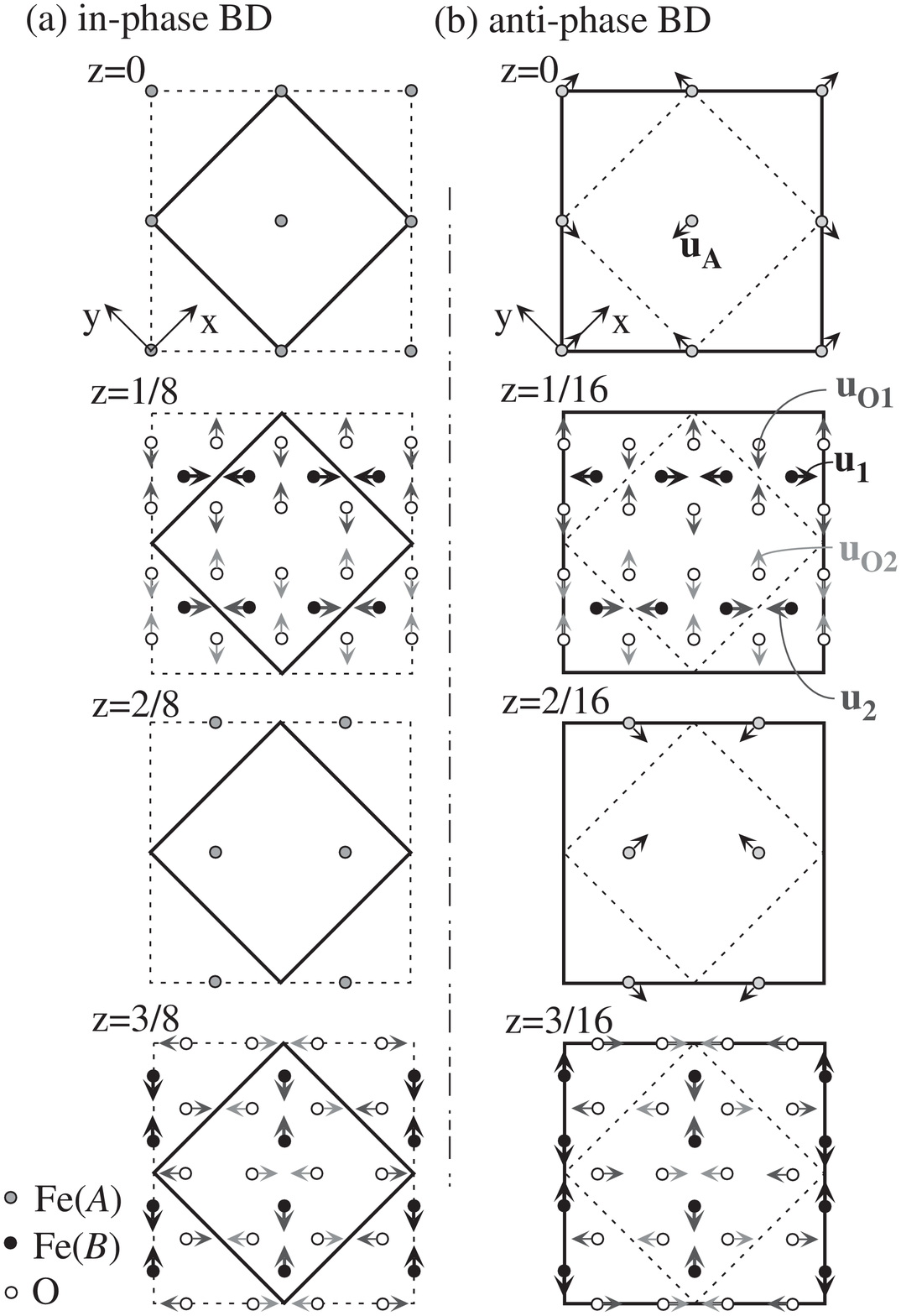,width=8cm}}
 \vspace{2mm}
 \caption{Schematic representation of the ionic displacements 
 in (a) the in-phase BD pattern and (b) the anti-phase BD
 pattern. 
 Only two Fe($B$)$_2$O$_3$ and Fe($A$) layers each are shown, 
 and the directions of displacement are represented by arrows. 
 The motions of the $A$ sites in (a) is not shown, 
 which are along the $z$ direction. 
 }
 \label{distortion}
\end{figure}

To see this competition, let us calculate 
a semi-phenomenological Landau-type free energy 
by taking into account of both 
in-phase and anti-phase BD states. 
We consider the amount of lattice distortion for the BD 
in the Fe($B$) ions to be uniform along each chains, 
denoted by $u_1$ and $u_2$, 
for alternative chains in the $xy$ planes. 
Then the in-phase BD is characterized by $u_1=u_2$
while the anti-phase BD is by $u_1=-u_2$, 
as shown in Fig.~\ref{distortion}(a) and (b), respectively. 
The motions of oxygens are approximated to be perpendicular to the chains 
in the $xy$ planes, 
parametrized by $u_{\rm O1}$ and $u_{\rm O2}$, 
for the chains with BD of $u_1$ and $u_2$, respectively. 
Within these approximations, the Fe($A$) ions 
becomes all crystallograhpically equivalent 
and we allow their motions in any direction, 
thus represented by the $x,y,$ and $z$ components of the displacement vector, 
$u_{{\rm A}x}$, $u_{{\rm A}y}$, and $u_{{\rm A}z}$, respectively 
(see Fig. \ref{distortion}).

Then the free energy per formula unit of Fe$_3$O$_4$, $\mathcal{F}$,  
can be computed, 
where there are 4 Fe($A$)-O bonds for $\mathcal{E}_{\rm Alat}$
and 12 Fe($B$)-O bonds for $\mathcal{E}_{\rm Blat}$, as 
%
%
%
\begin{eqnarray}
\mathcal{F} &=& \mathcal{F}_{\rm BD} + \mathcal{E}_{\rm Alat} 
  + \mathcal{E}_{\rm Blat}, \\
\mathcal{F}_{\rm BD} &=& \alpha(T-T_{\rm c})(u_1^2 + u_2^2), \\
\mathcal{E}_{\rm Alat} 
&=&
\frac{K_{\rm A}}{2} 
\Bigg\{ \left( \sqrt{u_{{\rm A}x}^2 
       + \left( \frac{1}{\sqrt{2}}+u_{{\rm A}y}-u_{\rm O1} \right)^2 
       + \left( \frac{1}{2} + u_{{\rm A}z} \right)^2 }
  - \sqrt{ \frac{3}{4} } \right)^2 
\nonumber\\ 
&\ & \hspace{10mm} 
       + \left( \sqrt{u_{{\rm A}x}^2 
       + \left( \frac{1}{\sqrt{2}}-u_{{\rm A}y}+u_{\rm O2} \right)^2 
       + \left( \frac{1}{2} + u_{{\rm A}z} \right)^2 }
  - \sqrt{ \frac{3}{4} } \right)^2 \nonumber\\
&\ & \hspace{10mm} 
　       + \left( \sqrt{
       \left( \frac{1}{\sqrt{2}}-u_{{\rm A}x}+u_{\rm O1} \right)^2 
       + u_{{\rm A}y}^2 
       + \left( \frac{1}{2} - u_{{\rm A}z} \right)^2 }
  - \sqrt{ \frac{3}{4} } \right)^2 \nonumber\\
&\ & \hspace{10mm} + \left( 
       \sqrt{\left( \frac{1}{\sqrt{2}}+u_{{\rm A}x}-u_{\rm O2} \right)^2 
       + u_{{\rm A}y}^2 
       + \left( \frac{1}{2} - u_{{\rm A}z} \right)^2 }
  - \sqrt{ \frac{3}{4} } \right)^2 \Bigg\}, \\
\mathcal{E}_{\rm Blat} 
&=& \frac{K_{\rm B}}{2} \Bigg\{
   2 \left( \sqrt{ \left( \frac{1}{\sqrt{2}}-u_1 \right)^2
          + \left( \frac{1}{ \sqrt{2} }+ u_{\rm O1} \right)^2 } -1 \right)^2 
   \nonumber\\
&\ & \hspace{10mm} +   2 \left( \sqrt{ \left( \frac{1}{\sqrt{2}}-u_2 \right)^2
          + \left( \frac{1}{ \sqrt{2} }+ u_{\rm O2} \right)^2 } -1 \right)^2  
   \nonumber\\
&\ & \hspace{10mm} 
 +   2 \left( \sqrt{ \left( \frac{1}{\sqrt{2}}+u_1 \right)^2
          + \left( \frac{1}{ \sqrt{2} }- u_{\rm O1} \right)^2 } -1 \right)^2  
   \nonumber\\
&\ & \hspace{10mm} 
+   2 \left( \sqrt{ \left( \frac{1}{\sqrt{2}}+u_2 \right)^2
          + \left( \frac{1}{ \sqrt{2} }- u_{\rm O2} \right)^2 } -1 \right)^2  
   \nonumber\\
&\ & \hspace{10mm} 
 + \left( \sqrt{1+\left(u_{\rm 1}-u_{\rm O1}\right)^2 } 
      -1 \right)^2 
+ \left( \sqrt{1+\left(u_{\rm 2}-u_{\rm O2}\right)^2 } 
      -1 \right)^2  \nonumber\\
&\ & \hspace{10mm} 
 + \left( \sqrt{1+\left(u_{\rm 1}+u_{\rm O2}\right)^2 } 
      -1 \right)^2 
+ \left( \sqrt{1+\left(u_{\rm 2}+u_{\rm O1}\right)^2 } 
      -1 \right)^2 \Bigg\},
\label{eqn:elastic}
\end{eqnarray}
%
\noindent 
where the lattice constant of the cubic unit cell is 
set to 4 so that the length of Fe($A$)-O and Fe($B$)-O bond 
without any distortion are 3/4 and 1, respectively. 
$\mathcal{F}_{\rm BD}$
describes the instability toward the BD state along each chain 
as the temperature decreases, 
discussed in the previous Section.  

\begin{figure}
 \centerline{\epsfig{file=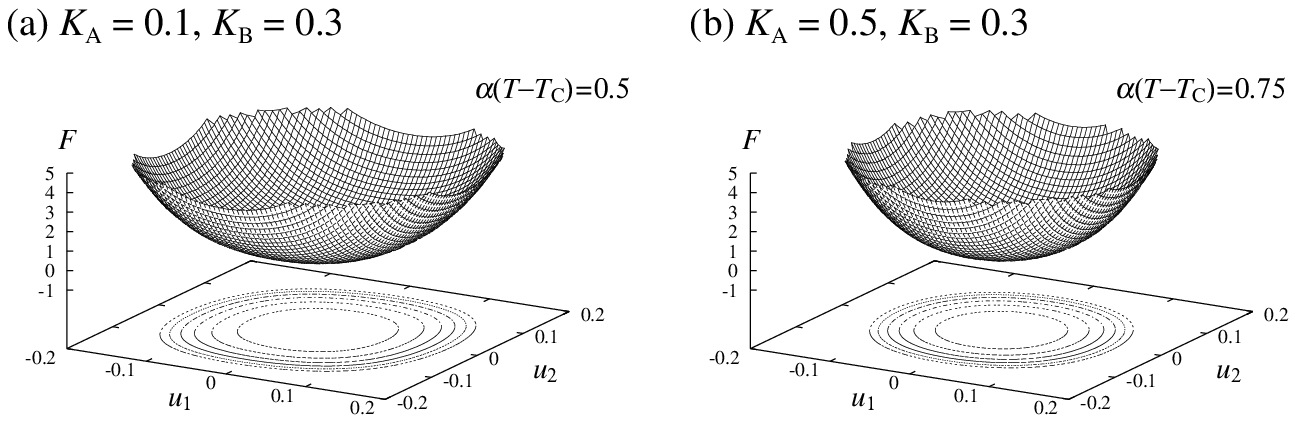,width=16cm}}
\vspace{-1cm}
 \centerline{\epsfig{file=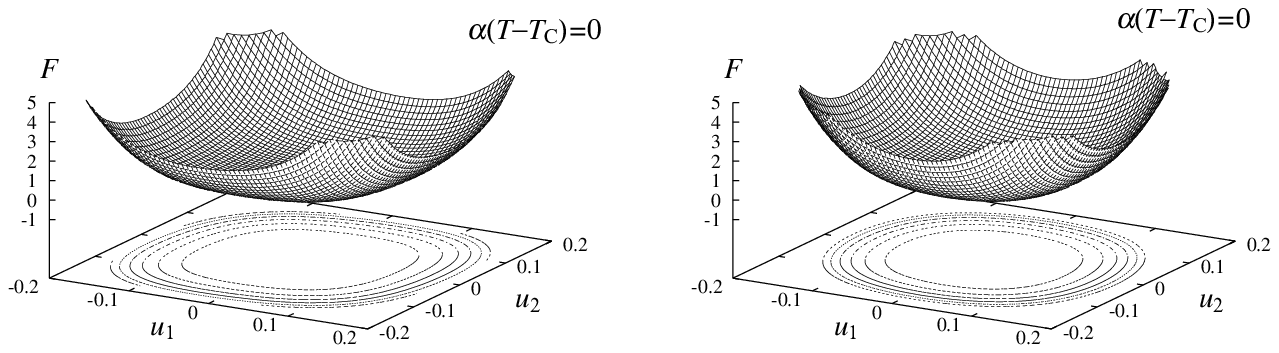,width=16cm}}
\vspace{-1cm}
 \centerline{\epsfig{file=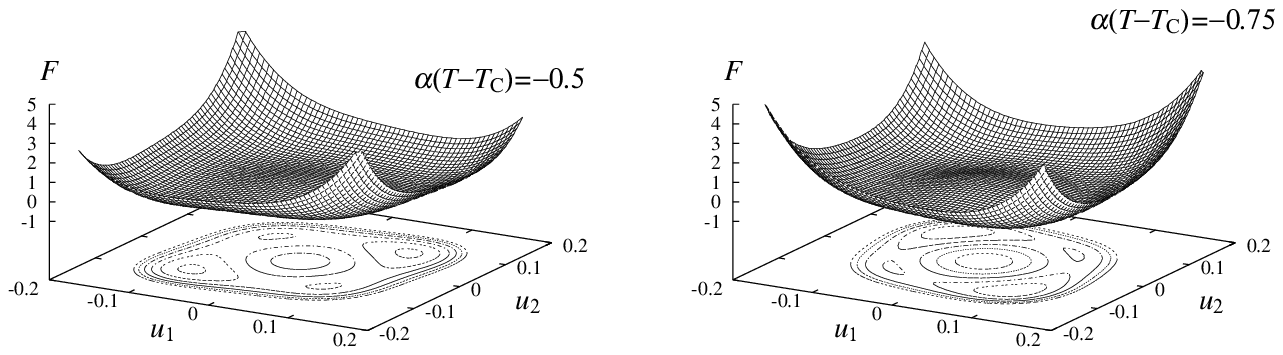,width=16cm}}
 \vspace{3mm}
 \caption{Calculated free energy, $\mathcal{F}$, 
 for (a) $K_{\rm A} = 0.1, K_{\rm B} = 0.3$ and 
(b) $K_{\rm A} = 0.5, K_{\rm B} = 0.3$ 
at several temperatures, 
as a function of the degree of dimerization for adjacent chains, ($u_1, u_2$). 
Note that the contour plot in the base is guide for eyes.
}
 \label{GLfig}
\end{figure}

\begin{figure}
 \centerline{\epsfig{file=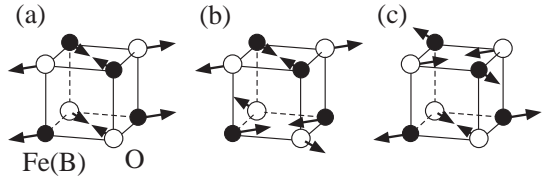,width=7.1cm}}
 \vspace{3mm}
 \caption{Schematic view of the lattice displacements 
 in the Fe($B$)-O cubes connecting adjacent Fe($B$)$_2$O$_4$
 layers when the BD state is stabilized. 
 In the in-phase BD pattern 
 only the displacement pattern $(a)$ is realized, 
 while in the anti-phase BD pattern 
 (a), (b) and (c) are realized with the ratio of $2:1:1$. 
 }
 \label{Bcubes}
\end{figure}

In Fig. \ref{GLfig}, $\mathcal{F}$
as a function of ($u_1$,$u_2$) is plotted
for the optimized positions of the Fe($A$) and O ions, 
obtained by minimizing it numerically by varying other variables, 
$u_{{\rm A}x}$, $u_{{\rm A}y}$, $u_{{\rm A}z}$, $u_{\rm O1}$, and 
$u_{\rm O2}$. 
It is plotted for two sets of elastic constants 
(a) $K_{\rm A} = 0.1, K_{\rm B} = 0.3$ and 
(b) $K_{\rm A} = 0.5, K_{\rm B} = 0.3$, 
at several temperatures. 
Above $T_{\rm c}$, 
the shape of the parabolic curvature gets deeper 
as $T_{\rm c}$ is approached, 
almost symmetrically along $u_1=u_2$ and $u_1=-u_2$ 
suggesting that the fluctuations of both in-phase and anti-phase 
BD states develop. 
Below $T_{\rm c}$, 
for the parameters in Fig. \ref{GLfig}(a),
the minimum of the free energy appears 
along $u_1=u_2$, that is, the in-phase BD state 
is stabilized, 
while in Fig. \ref{GLfig}(b) the mimimum is along $u_1=-u_2$ 
which means that the anti-phase BD state is realized. 
Note that the phase transition is a second order one. 
%
%
%
The stability of these BD patterns
can be understood as follows. 
In the presence of the in-phase BD pattern, 
all the Fe($B$)$_4$O$_4$ cubes~\cite{note3}, 
connecting the Fe($B$)$_2$O$_4$ $xy$ planes  
show ionic displacements 
as shown in Fig.~\ref{Bcubes}(a), 
where all the interlayer Fe($B$)-O pairs 
move respectively in the same direction, 
providing small $\Delta u_{\rm B-O}$'s. 
Thus the lattice elastic energy of the Fe($B$)-O bonds, 
${\mathcal E}_{\rm Blat}$, 
should be the lowest for the in-phase BD states with such configuration, 
compared to other BD patterns 
containing cubes 
as shown in Figs.~\ref{Bcubes}(b) and \ref{Bcubes}(c), 
with both Fe($B$) pairs forming dimers
or both being inter-dimer Fe($B$) pairs, respectively.
For example, the anti-phase BD 
shows all of these three kinds of cubes, 
realized with the ratio of $2:1:1$, in order, 
thus costing rather higher ${\mathcal E}_{\rm Blat}$ 
than in-phase BD. 
\begin{figure}
 \centerline{\epsfig{file=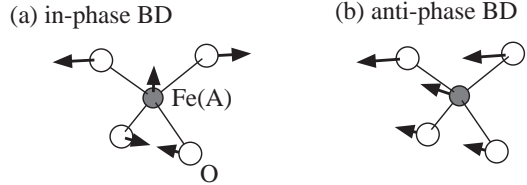,width=7.1cm}}
 \caption{Schematic view of the displacements 
 in the Fe($A$)-O tetrahedra 
 for (a) the in-phase BD pattern and (b) the anti-phase BD pattern.}
 \label{Asites}
\end{figure}
On the other hand, 
the Fe($A$)-O tetrahedra would be quite deformed 
in the presence of the in-phase BD, 
which is shown in Fig.~\ref{Asites}(a), 
providing large lattice elastic energy  
of the Fe($A$)-O bonds, 
${\mathcal E}_{\rm Alat}$, 
while the anti-phase BD pattern
make the O ions surrounding the Fe($A$) ions 
showing displacements 
as shown in Fig.~\ref{Asites}(b), 
where the cost in ${\mathcal E}_{\rm Alat}$ will be rather smaller. 
This is because the motion of the oxygens 
are in the same direction for the latter pattern 
so that the Fe($A$) ions can adjust their position to lower 
$\Delta u_{\rm A-O}$, 
as can be seen in Fig.~\ref{Asites}(b).
To summarize, one can say that ${\mathcal E}_{\rm Blat}$ favors the in-phase 
BD state while ${\mathcal E}_{\rm Alat}$ favors the anti-phase one. 

These naive discussions are consistent with 
the above calculation of the Landau-type free energy. 
There,
we have observed that a competition arises between 
the in-phase and anti-phase BD states, 
stabilized for $K_{\rm A} \lsim K_{\rm B}$ 
and for $K_{\rm A} \gsim K_{\rm B}$, respectively. 
We propose that the latter is the case in magnetite, 
since in general, 
large deformation is hardly realized 
in the tetrahedra of oxygen surrounding the $A$ sites 
of the spinel structure, 
which may be due to its tight packing 
compared to the $B$ sites~\cite{Verway3}. 
The importance of such ionic displacement of 
the O ions surrounding the $A$ sites 
in determining the electronic properties in the ground state 
has recently also been pointed out in another spinel compound 
AlV$_2$O$_4$~\cite{Katsufuji}. 
This anti-phase BD state provides the correct unit cell 
determined experimentally~\cite{Iizumi}, 
as will be discussed later. 

\section{comparison with experiments}\label{secexp}

\begin{figure}
 \centerline{\epsfig{file=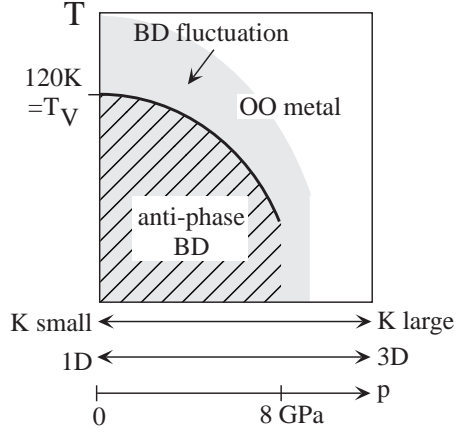,width=6.5cm}}
 \vspace{1mm}
 \caption{Schematic phase diagram for magnetite Fe$_3$O$_4$. 
 The OO and BD denotes the orbital-ordered and bond dimerized states, 
 respectively. 
 The vertical axis is the temperature, $T$, 
 while the horizontal axis can be interpreted 
 either as the pressure, $p$, the ``rigidness'' of the lattice 
 represented by the lattice constant, $K$, or 
 the effective dimensionality. 
 The bold line is the Verway transition temperature, $T_{\rm V}$, 
 varying as a function of such parameters. } 
 \label{phasediagram2}
\end{figure}

Based on the discussions above, 
our proposal for 
the physical properties in magnetite Fe$_3$O$_4$ can be 
summarized in Fig.~\ref{phasediagram2}, 
which has the following new features: 
the existence of OO in the $B$ sites 
at all temperature range below room temperature, 
the BD fluctuation (both the in-phase and the anti-phase BD's) 
above the Verway transition temperarure, $T_{\rm V}$, 
and the anti-phase BD stabilized in the insulating phase 
as a consequence of compensation 
with the lattice elastic energy of the whole system. 
These are compared with the known experimental facts 
in the following. 

Below the Verway transition temperature,
the existence of ferro-OO is supported with 
dielectric measurements showing large anisotropy~\cite{dielectric}.
Above $T_{\rm V}$, in contrast, crystal structure analysises 
show no evidence of the lowered symmetry from the cubic one, 
apparently contradicting with 
our prediction of the ferro-OO even in the metallic phase.  
However, 
the symmetry in the electronic structure 
has been pointed out 
to be lowered than the cubic one~\cite{Siratori} 
based on magneto-crystalline anisotropy measurements~\cite{magneto} 
and in the recent resonant X-ray scattering at room
temperature~\cite{Garcia}, 
consistent with the OO state.

As for the BD, 
there is a strong support from neutron scattering measurement 
by Shapiro~{\it et~al.}~\cite{Shapiro}, 
which is the observation of {\it 1D correlation} above $T_{\rm V}$, 
not explained by the CO but consistent with 
the present BD picture. 
This 1D nature of the BD transition can 
naturally provide an explanation for 
the critical fluctuation above $T_{\rm V}$ 
observed in large temperature range up to room temperature, 
such as in the pseudogap structure of the optical conductivity~\cite{Park} 
and in neutron scattering measurements 
where it has actually been realized that 
the atomic displacement is playing an important role~\cite{Fujii}. 
Moreover, our scenario that the Verway transition 
is a phase transition from OO metal to the coexistent state of OO and BD 
explains the entropy change of $\Delta S = R \log 2$ per mole of Fe$_3$O$_4$ 
estimated from specific heat measurements~\cite{Shepherd}, 
arising from the degree of freedom for dimerization. 
This excludes 
the order-disorder-type CO picture 
of Verway~\cite{Verway2} predicting $\Delta S = 2R \log 2$ 
as well as 
simultaneous OO and BD transition 
with $\Delta S =2R \log 3$ from the orbital degree of freedom.

The crystal structure in the presence of anti-phase BD 
provides the correct unit cell size 
below $T_{\rm V}$ in the actual compound, 
namely, $\sqrt{2} \times \sqrt{2} \times 2$ of 
the cubic unit cell above $T_{\rm V}$ 
shown in Fig.~\ref{structure}~\cite{Iizumi}, 
which is also supported 
by the double period modulation along the $z$ axis 
observed in electron diffraction~\cite{electron} 
and in neutron scattering~\cite{Samuelson}. 
Furthermore, the diffusive spot above $T_{\rm V}$
of the corresponding wave vector $(0,0,1/2)$, 
observed in a neutron scattering measurement 
showing a divergent behavior toward $T_{\rm V}$~\cite{Fujii} 
can be assigned to the anti-phase BD fluctuation 
developping above $T_{\rm V}$ 
seen in the calculation of Sec.~\ref{secBDpattern}. 
On the other hand, 
the analysis of Yamada {\it et~al.}~\cite{Yamada} 
of their neutron scattering data above $T_{\rm V}$
apparently seems to correspond to the in-phase one, 
where the ionic motions of the Fe($B$)-O cubes 
are proposed to be as in Fig. \ref{Bcubes}(a). 
This might be the observation of the 
fluctuation of the in-phase BD mode, 
which is also seen in our calculation 
though a long range order of this mode is not achieved.  
We have seen that such competition arises 
due to the existence of two kinds of elastic energies, 
${\mathcal E}_{\rm Alat}$ and ${\mathcal E}_{\rm Blat}$, 
which may be consistent with the experimental findings 
by Shapiro {\it et al.}
that ``two types of interacting dynamical variables 
are taking part in the neutron scattering''~\cite{Shapiro}. 

The observed pressure induced metallic behavior~\cite{Todo} 
can be naturally understood 
since the pressure should effectively increase 
the value of $K$ in Eq.~(\ref{eqn:H_P-H}) 
(or $K_{\rm A}$ and $K_{\rm B}$ in Eqs.~(\ref{Alat}) and (\ref{Blat})~)
making the lattice more ``rigid'', 
and/or 
increase the three-dimensionality of the system, 
both expected to destabilize the BD state, 
which is represented in Fig.~\ref{phasediagram2}. 
This is analogous to the case of spin-Peierls transition, 
the mapped state of the BD state as mentioned in Sec.~\ref{subsec1D}, 
known to be destroyed by enhancement in 
the electron-lattice coupling constant 
and/or the three-dimensionality~\cite{Inagaki}. 
We point out the possibility that the three-dimensional BD pattern 
may be changed under different environments such as pressure and temperature, 
since there are two competing phases, the in-phase and the anti-phase BD 
states, as seen in Sec.~\ref{secBDpattern}. 
Search for such transition between these phases 
under uniaxial pressure is also interesting, 
since these states are highly anisotropic.  
Another possibility is that incommensurate phases 
may be stabilized as a consequence of such competition, 
as observed 
in dielectrics [N(CH$_3$)$_4$]$_2M$Cl$_4$~\cite{FujiiYuuden}
as well as in CO systems such as perovskite Ni oxides~\cite{nickelate}
and NaV$_2$O$_5$~\cite{Ohwada},
due to competing different states. 

One discrepancy between our theory and the experimental 
facts is the order of the phase transition: 
experimentally it is first order while it is 
second order in our calculations in Secs.~\ref{secBsites}
and \ref{secBDpattern}. 
This discrepancy may be due to the simplified approximation 
we made for the strucural change 
as discussed in Sec.~\ref{secBDpattern}. 
In the actual compound, 
the lattice distortion will be more complicated 
than in our calculation, 
represented as in Fig.~\ref{distortion}, 
e.g., the lattice distortion of the $B$ sites 
along the chain should 
take the period of four sites which is contained in the unit cell, 
as $(u_a, -u_b, u_c, -u_d,u_a,-u_b,...)$, 
rather than $(u,-u,u,-u,...)$ as in our calculation. 
This should make the crystal symmetry of the BD insulating phase 
lower than that in our calculation, 
which may lead to a first order phase transition~\cite{Landau}.

Finally, it is noted that the recent findings of 
possible charge disproportionation in the $B$ sites 
by X-ray anomalous scattering~\cite{Toyoda} 
does not contradict with our proposal here. 
The anti-phase BD state gives rise to 
crystallographically independent Fe($B$)~ions, 
which should make the charge density on the $B$~sites different 
where the amount of charge disproportionation 
will be not so large 
but can be detectable in a sensitive probe such as  
X-ray anomalous scattering measurements. 


\section{conclusion}\label{secconcl}

In conclusion, 
we have theoretically proposed 
a new model for the mechanism of 
the Verway transition in magnetite Fe$_3$O$_4$. 
It is not a charge ordering transition 
as has been believed for more than a half century, 
but a bond dimerization induced by the Peierls instability 
in a one-dimensional state as a consequence of the ferro-orbital ordering 
due to strong electronic correlation. 
The actual bond dimerization pattern 
is predicted to be the anti-phase one, 
based on discussions on the lattice elastic energy of the system. 
This model seems to be able to provide 
the explanation for the long lasting mystery 
of the Verway transition in this compound. 

\vspace{0.5cm}

\acknowledgements

We thank Y. Fujii, T. Katsufuji, N. M\^{o}ri, N. Nagaosa, Y. Nakao, K. Ohwada, 
H. Takagi and S. Todo for valuable discussions and suggestions.
We also thank T. Katsufuji and T. Toyoda for providing us 
Refs. \ref{Katsufuji} and \ref{Toyoda}, respectively, 
and A. Himeda and T. Koretune for technical supports 
on numerical calculations. 
This work is supported by the Grant-in-Aid from 
Ministry of Education, Science, Sports and Culture of Japan.


\begin{thebibliography}{99}
\bibitem{note}The $t_{2g}$-orbital level in the spinel
	structure splits into $e_g$ and $a_{1g}$ symmetry combinations,
        though, band structure calculation shows that 
	the octohedral component in the crystal field 
        at the $B$ site in Fe$_3$O$_4$ is so strong that the splitting 
	is negligible~\cite{Zhang}.
\bibitem{note2}We have repeated the calculation of 
	Mishra {\it et al.}~\cite{Mishra} and the value we obtained 
	for $V_{\rm c}$ was 0.29~eV, smaller than theirs. 
	The difference should be due to the larger number of 
	mesh in our calculation taken for the {\it k}-integration,
	though it is not important for the discussion here.
\bibitem{note3}Strictly saying, the Fe(B)-O network is formed not in cubes 
	but in parallelepipeds, 
	though it is approximately cubes so the we call them so in
	this paper for simplicity.
\end{thebibliography}
\end{document}